\documentclass{article}
\usepackage[utf8]{inputenc}
\usepackage{graphicx}
\usepackage{epstopdf}
\usepackage{amsmath}
\usepackage{subfigure}
\usepackage{amssymb}
\usepackage{mdframed}
\usepackage{hyperref}
\usepackage{dsfont}
\usepackage{tcolorbox}
\usepackage{lineno}
\usepackage{tikz}
\usepackage{authblk}

\newcommand{\cat}[1]{\textit{#1}}

\begin{document}

\title{Diagnosing the performance of human mobility models at small spatial scales using volunteered geographic information}

\author[1]{Chico Q.\ Camargo}
\author[1]{Jonathan Bright}
\author[1,2]{Scott A.\ Hale}
\affil[1]{Oxford Internet Institute, University of Oxford, UK}
\affil[2]{Alan Turing Institute, London, UK}

\maketitle

\begin{abstract}
{\bf
Accurate modelling of local population movement patterns is a core contemporary concern for urban policymakers, affecting both the short term deployment of public transport resources and the longer term planning of transport infrastructure. Yet, while macro-level population movement models (such as the gravity and radiation models) are well developed, micro-level alternatives are in much shorter supply, with most macro-models known to perform badly in smaller geographic confines. In this paper we take a first step to remedying this deficit, by leveraging two novel datasets to analyse where and why macro-level models of human mobility break down at small scales. In particular, we use an anonymised aggregate dataset from a major mobility app and combine this with freely available data from OpenStreetMap concerning land-use composition of different areas around the county of Oxfordshire in the United Kingdom. We show where different models fail, and make the case for a new modelling strategy which moves beyond rough heuristics such as distance and population size towards a detailed, granular understanding of the opportunities presented in different areas of the city. 
}
\end{abstract}

\section{Introduction}

Predicting human mobility is important for urban planning, traffic control, and for the general management of a city. This question has been addressed by a long tradition of mathematical models of human mobility, including the gravity model, dating back to Zipf\cite{zipf1946p} and Carey\cite{carey1867principles}, Stouffer's 1940 intervening opportunities model \cite{stouffer1940intervening,schneider1959gravity}, as well as the celebrated parameter-free radiation model, by Simini et al.~\cite{simini2012universal}

Despite its many successes at predicting large-scale movement, the radiation model is known to perform poorly in predicting traffic at small spatial scales. The modified radiation model by Yang et al. meant to address that issue by introducing a scaling parameter $\alpha$ representing the influence of small spatial scales in human mobility \cite{yang2014limits}. This extra parameter, as described in the original paper, would be a way to address the separation between population density and trip attraction rates that happens at small scales \cite{yang2014limits}.

This problem of poor mobility predictions at small spatial scales has also been addressed by considering the variation in the accessibility of different sites~\cite{Piovani2018} as well as features like the topology of urban spaces \cite{brelsford2018toward}. While there is no single definition of accessibility---it is often defined depending on its particular application~\cite{geurs2001accessibility,Bertolini2005,Vandenbulcke2009,Batty2009}---it typically relates the opportunity of accessing a specific location to the cost of travel, which is essential to urban planning~\cite{Batty2009}.

It makes sense to include fine-grained variables when predicting mobility at small scales, since the smaller the scale of a system the harder it becomes for its variables to average out. For example, the travel time between different neighbourhoods in a city depends not only on their distance, but on their accessibility, while at a national scale one can roughly assume all cities are equally accessible since there are major roads connecting to most regions of a country and other accessibility differences might average out. Note that it remains a very rough assumption, but the point is that it becomes a safer assumption at larger scales. This behaviour is well-known in the physical sciences, where larger systems also tend to be more predictable than smaller systems, due to asymptotic effects such as the law of large numbers and the central limit theorem \cite{kotelenez1986law}. Take, for example, the diffusion of ink in a glass of water: while the motion of any small group of ink molecules might show random fluctuations, once one considers the whole ink solute, the randomness essentially disappears as the solute gradually dissolves from its high-concentration areas to low-concentration areas. Thus, in chemistry, diffusion is described as a deterministic macroscopic phenomenon, despite its microscopic random nature~\cite{berg1993random}. The same applies to human mobility: at large spatial scales, the differences between different regions of a city average out. At small spatial scales, more data is required.

Mobility models traditionally use demographic and geographic data as input, being typically limited to the population of the zones of interest and the distance between them. In the last decades, new sources of data have become available. In addition to traffic sensors that can estimate in real time the volume of vehicles flowing through the streets, many cities are now home to a variety of sensors of traffic congestion, urban noise levels, air quality, and other variables such as water and energy usage. The technology used in urban sensing also includes video surveillance powered by image recognition tools allowing the detection and classification of different types of vehicles as well as cyclists and pedestrians~\cite{Cordts2016Cityscapes}.

In fact, much of this sensing technology comes not in the form of dedicated devices located in strategic spots on roads and buildings, but rather integrated to mobile devices. The ever-growing number of mobile applications for urban routing, ride-sharing and sharing of geolocated information in social media, aided by the ubiquity of mobile phones, have turned these devices into ``floating sensors''\cite{jawhar2010inter}, a valuable source of detailed geographic data. Recent studies in urban mobility have tapped into the potential of these new data sources, achieving successful traffic prediction from mobile phone and social media data alone or in combination with demographic and sensor data \cite{mcneill2017estimating,WazePaper,Louail2015,Barbosa2018}.

A third source of data is land use data. There is a vast literature on the links between land use and transport \cite{wegener2004land,Lenormand2015,Louail2015}, often classifying land use types at highly aggregate levels, such as residential, commercial or industrial areas. Historically, compiling land usage datasets has been an expensive and time consuming task \cite{liu2012urban}. The volunteered geographic information site OpenStreetMap, launched in 2004, can be seen as an alternative solution to this problem: rather than being compiled by a single person or team, OpenStreetMap is the outcome of a Wikipedia-like collaborative editing process that produces a free, open and detailed world map, including details such as the location and classification of different types of residential and commercial locations. The accuracy and completeness of OpenStreetMap coverage has been assessed in several studies \cite{haklay2010good,girres2010quality,zielstra2010comparative,helbich2012comparative,mashhadi2015impact,arsanjani2015quality,senaratne2017review,bright2018geodemographic,bright2018openstreetmap}, yielding positive but cautious results, particularly about road networks. 

The diversity of new data sources for traffic prediction described above becomes a promising resource, once one considers that human mobility at smaller scales is less likely to behave according to laws as simple as the ones proposed by gravity or radiation models. With that in mind, rather than using these physics-based models of traffic prediction, one can use large amounts of data to train machine learning models that hold little or no \textit{a priori} assumptions about the system. This data-driven approach has proved quite successful in traffic prediction \cite{hasan2013spatiotemporal,iqbal2014development,BenZion2018, WazePaper}. The problem often raised about this approach is that it does not always add explanatory power, as more complex machine learning models are known to be ``black boxes'' which work quite well but are quite challenging to explain or interpret \cite{ribeiro2016should,marcus2018deep}. Additionally these approaches have a strong historical bias, which could result in poor predictions for areas undergoing rapid development or change \cite{hand2006classifier,vzliobaite2010learning}.

An alternative to complex, non-interpretable machine learning models are human mobility models that take into account very fine-grained details such as the topology of a city \cite{brelsford2018toward} or the accessibility of different sites \cite{Piovani2018,Batty2009}. In contrast to purely data-driven approaches, these models often bring with them very mechanistic assumptions about mobility, which makes it possible to evaluate such assumptions against each other. The disadvantage is that they often require knowing quite a lot about a city and use data that is not always available or easy to obtain. While there is an large increase in the availability of urban data, such data remain unevenly distributed: there is a lot of data about major cities such as London, New York, Paris, and Tokyo, but much less data about smaller or poorer places \cite{zielstra2010comparative,haklay2010good,thebault2018geographic}.

This suggests that the failure of small-scale traffic prediction will be dominant in smaller cities, which often also have smaller budgets. 

Dedicated sensors are not always an affordable choice for local governments, and even social media data might be too scarce or expensive to obtain. In addition, ride-sharing services and navigation applications are not available in smaller cities. All these considerations make a point for models using open data, such as the OpenStreetMap described above, as an affordable way to improve traffic prediction at small spatial scales.

In this paper, we use OpenStreetMap data in combination with public demographic data to explore how different human mobility models perform at small spatial scales, comparing models against two months of mobility data in the county of Oxfordshire, UK. First, we describe the dataset and the eight classes of models used in this study. We then fit each class of models to the data, and show that most models fail at predicting trips at this spatial scale.

Next, we a propose a series of standard ways to modify the models, such as using travel time instead of distance and adding a multiplicative term to correct the underestimated urban-to-urban trip volumes. We show that they do not solve the inaccuracies in the models. We argue, in agreement with the literature, that solving this problem requires more than using demographic and distance data, such as data about what is present in the origin and destination zones, for which we use OpenStreetMap data in this paper. 

In the end, we discuss the applicability of our method to different locations, and its relation to traffic modelling at different spatial scales.

\section{Case study: Oxfordshire, UK}

\begin{figure*}[htp]
\begin{center}
\includegraphics[width=0.9\linewidth]{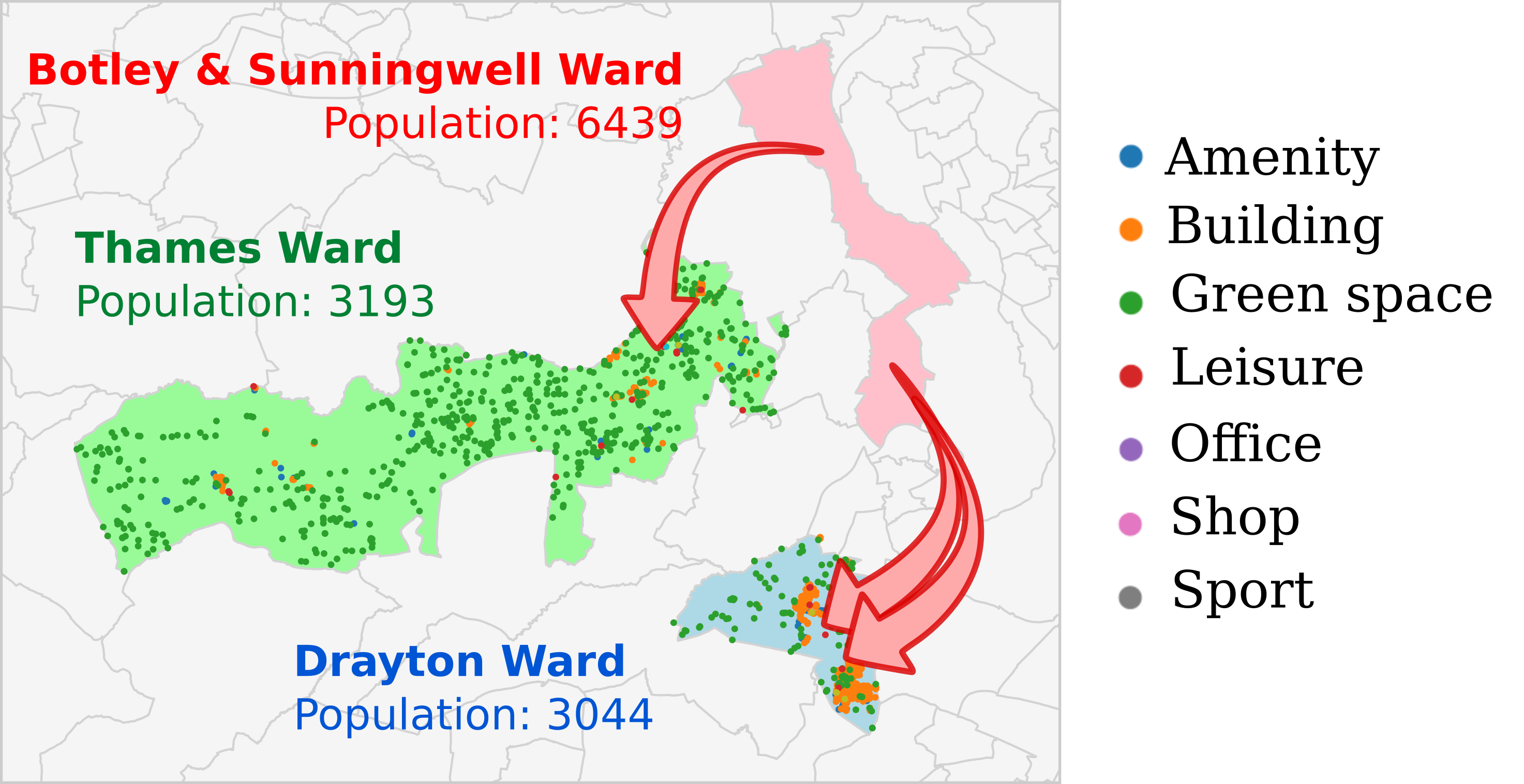}
\end{center}
\caption{\label{fig:wards_illustration}
Illustration of the difference between the trip volumes between different zones in Oxfordshire. In the figure, both destinations (Thames Ward and Drayton Ward) have similar resident and working populations, and both are at approximately 11 km from Botley \& Sunningwell Ward. However, there are about twice as many trips from Botley to Drayton as there are from Botley to Thames. This might be explained by the difference in the composition of both wards, shown by their points of interest on OpenStreetMap: Drayton Ward is remarkably more urban compared to Thames Ward.}
\end{figure*}

Our case study examines Oxfordshire, a county in South East England, with an area of 2,605 km$^2$, and a total population of approximately 680,000 inhabitants. Counties in the United Kingdom are divided into electoral wards, or simply wards, which are usually named after neighbourhoods, parishes and other geographical marks. The names and borders of different wards are defined by the local government, and may change over time. We obtained ward-level mid-year population estimates for the 112 Oxfordshire wards, as defined in April 2016, from the Office of National Statistics. Shapefiles describing the border of all Oxfordshire wards were downloaded from the Digimap mapping data service \cite{digimap}.

For our OpenStreetMap data, we downloaded the \cat{points of interest} from the OpenStreetMap database~\cite{OpenStreetMap} in November 2017, which are geolocated points tagged with indications of the kind of land use found in that location. We downloaded 1,071,877 points of interest within Oxfordshire, and selected points tagged as amenity, building, land use (i.e., green spaces), leisure, office, shop, and sport. All points of interest also contain `minor' subtags, such as \cat{pub}, \cat{restaurant} or \cat{cafe}, all of which might have the same major tag of \cat{amenity}. In this paper, we chose to not use points of interest related to the transport network, such as \cat{railway} or \cat{street}, due to their poor coverage. We also did not consider categories with less than a hundred points of interest in Oxfordshire.

We used anonymised and aggregated GPS mobile phone data provided by a major smartphone operating system.

Similar data have previously been validated and successfully used in San Francisco~\cite{sana2017using} and Amsterdam~\cite{knoop2018empirical}. The data contains estimated trip volumes for origin--destination pairs of wards in Oxford for the January and February 2017 in hourly increments. We subset the data and only use trips inferred to be made by vehicle (and not walking or cycling) and trips on weekdays made between 7am and 12pm (noon).\footnote{We also experimented using the whole day and including weekend trips, but the overall results were qualitatively similar.}
We calculate the centroid of each ward and compute the geodesic distance between all centroid pairs.
Finally, we obtained the travel time between the centroids of all $112 \times 112$ origin--destination pairs of wards using the Google Distance Matrix API, part of the Google Maps Platform \cite{GoogleMapsAPI}.

\section{Human mobility models}\label{sec:models}

Here we explore eight classes of human mobility models, and analyse how well they perform in predicting traffic between different zones within the county of Oxfordshire. We use four variations on the gravity model \cite{zipf1946p,carey1867principles}. In its simplest form, the traffic volume $T_{ij}$ from zone $i$ to zone $j$ can be expressed as:

\begin{equation}\label{eq:gravitymodel}
T_{ij} = A \: n_i^\alpha n_j^\beta f(d_{ij})
\end{equation}

\noindent where $n_i$ and $n_j$ represent the population of zones $i$ and $j$, $A$ is a normalisation constant, and $f(d_{ij})$ represents a weighting cost function indicating the relation between the number of commuters $T_{ij}$ and the commuting cost $d_{ij}$, which is typically the distance or travel time between zones $i$ and $j$. This dependency is typically modelled $f(d_{ij}) = d_{ij}^{-\gamma}$ or $f(d_{ij}) = e^{- \gamma d_{ij}}$. Finally, $\alpha$, $\beta$ and $\gamma$ are tunable parameters. The gravity model is often fit by constraining the total traffic going in ($\sum_i T_{ij}$) and out ($\sum_j T_{ij}$) of every zone to have the same values as in the ground truth data, in what is called the doubly-constrained gravity model \cite{wilson1969use}. A common simpler version of the model, more similar to the Newtonian law of gravity, is found by setting the exponents of $n_i$ and $n_j$ to $\alpha = 1$ and $\beta = 1$ \cite{wilson2013entropy}. In this paper, we use four versions of the gravity model, given by the four permutations of the power function and the exponential function for $f(d_{ij})$ and whether $\alpha$ and $\beta$ are set to 1 or whether they are free to vary.

In addition to the gravity models discussed above, our analysis also includes the parameter-free radiation model, by Simini et al.~\cite{simini2012universal}. In this model, the $s_{ij}$ indicates the number of opportunities between wards, measured as the total population within a circle of radius $r = d_{ij}$, where $d_{ij}$ is the distance between wards $i$ and $j$. This $s_{ij}$ term, sometimes described as the number of intervening opportunities~\cite{schneider1959gravity}, can be seen as a proxy for the number of alternative places where a person from ward $i$ could go in less time than what would take them to reach ward $j$. This name is no coincidence, as the intervening opportunities model can be seen as the basis for the radiation model~\cite{simini2012universal}. The radiation model is defined as:

\begin{equation}\label{eq:radiationmodel}
T_{ij} = T_i \frac{n_i n_j}{ (n_i+s_{ij})(n_i + n_j + s_{ij}) }
\end{equation}

\noindent We also include Yang et al.'s one-parameter radiation model \cite{yang2014limits}:

\begin{equation}\label{eq:radiationalpha}
T_{ij} = T_i \frac{ [(n_i + n_j + s_{ij})^\alpha - (n_i + s_{ij})^\alpha] (n_i^\alpha + 1) }{ [(n_i+s_{ij})^\alpha + 1][(n_i + n_j + s_{ij})^\alpha + 1] }
\end{equation}

In this model, $\alpha$ measures the influence of small spatial scales in human mobility, by changing the coupling between population density and trip attraction rates at small distances~\cite{yang2014limits}. We include two versions of the modified radiation model: one with with the parameter $\alpha$ unconstrained, another with its value defined fit according to the average size of Oxfordshire wards, according to the formula in \cite{yang2014limits} of $\alpha = (\: l / 36[\text{km}] \:)^{1.33}$, where $l$ stands for the average size of Oxfordshire wards.

Finally, we also include the intervening opportunities model, as formulated by Schneider \cite{schneider1959gravity}, one of the original inspirations of the radiation model \cite{simini2012universal}. The intervening opportunities model is defined as:

\begin{equation}\label{eq:inter_oppo}
T_{ij} = e^{-\gamma s_{ij}} - e^{-\gamma( s_{ij} + n_j)}
\end{equation}

The performance of all eight models with the Oxfordshire traffic data is presented in Figure~\ref{fig:raw_models}.

\section{Mobility at small spatial scales}

Figure~\ref{fig:wards_illustration} illustrates a typical pair of trips within the Oxfordshire dataset. The figure shows three wards in Oxfordshire: Botley \& Sunningwell Ward, Thames Ward, and Drayton Ward. The latter two have similar resident and working populations, and both are at approximately 11 km from Botley \& Sunningwell Ward (distance measured from their centroids). The number of opportunities between wards, expressed as $s_{ij}$ in the models in section~\ref{sec:models}, is the only variable in the original models that varies between the Botley--Drayton trip and the Botley--Thames origin-destination pairs: it is $60\%$ higher for the Botley--Drayton trip, which would suggest this origin-destination pair would have a lower volume of traffic than the Botley--Thames pair. The data, however, shows the opposite trend: the traffic volume for the Botley--Drayton trip is approximately twice the volume for the Botley--Thames trip.

Out of many factors that could possibly explain the higher trip volume for the Botley--Drayton trip, one simple explanation jumps out when plotting the OpenStreetMap points of interest for the Drayton and Thames Wards. The Thames Ward is filled with points of interest classified as \cat{green space}, which indicate a large amount of green spaces. Although Drayton Ward also shows a large number of points marked as \cat{green space}, it also has more points classified as \cat{building}, indicating urban locations. These urban centres correspond to the villages of Drayton and Milton, which together account for most of the population of Drayton Ward. On the other hand, Thames Ward is composed of a handful of smaller villages and hamlets, which altogether make a much less urban environment despite a similar total population.

\section{Model performance at small spatial scales}

\begin{figure*}[htp]
\includegraphics[width=1.0\linewidth]{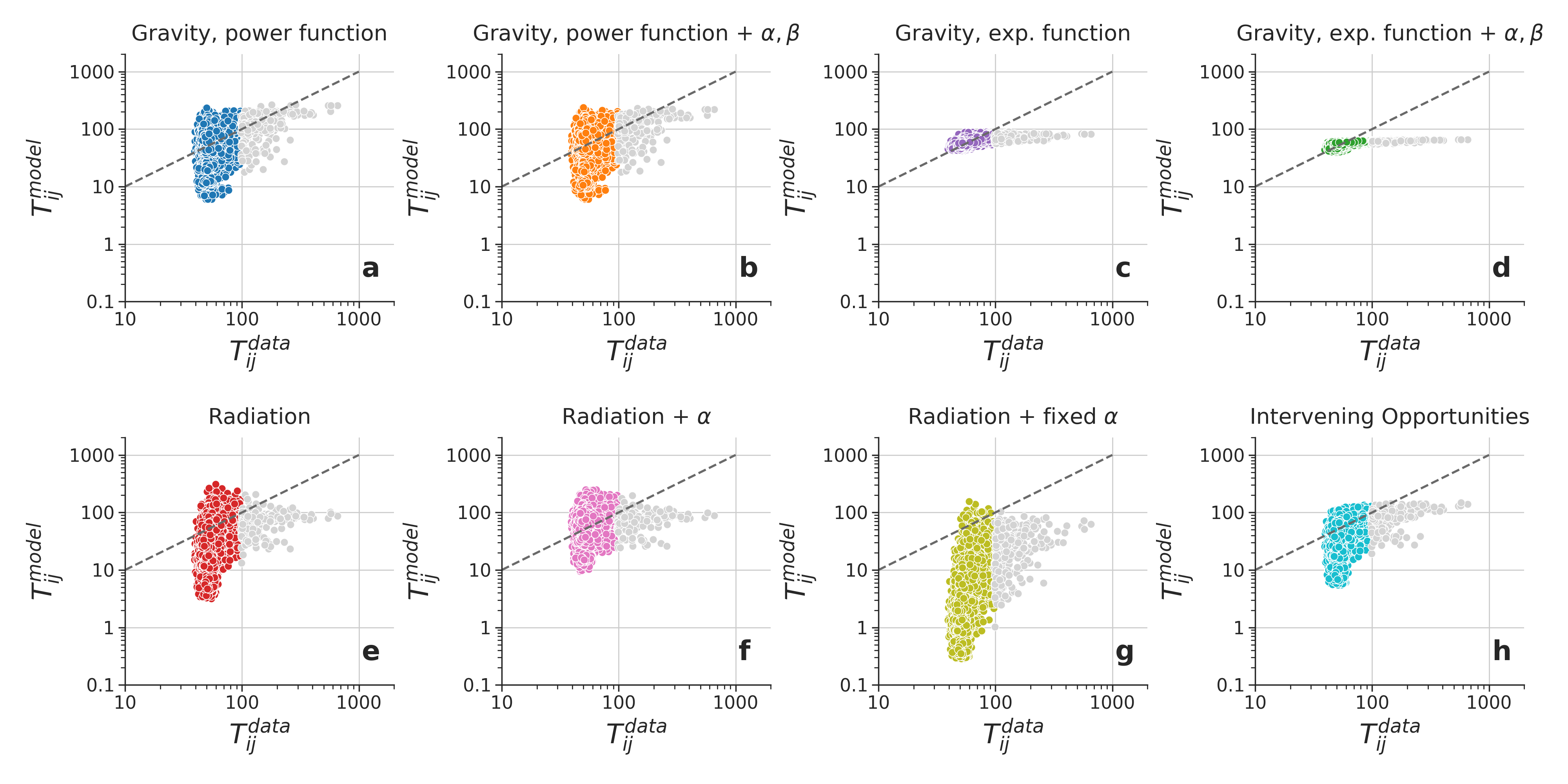}
\caption{\label{fig:raw_models}
\textbf{Current mobility models do not perform well at small spatial scales.} 
Panels \textbf{(a)} to \textbf{(h)} show the predicted trip volume $T_{ij}^{\text{model}}$ versus the trip volume $T_{ij}^{\text{data}}$ according to the mobility data, both plotted in logarithmic scale, for all models described in section~\ref{sec:models}.
All models show a poor fit to the data, especially for high-volume trips. Grey dots indicate the $2\%$ trips with the highest volume, which correspond to $14\%$ of all traffic volume, and dashed lines indicate the $y=x$ identity line.
}
\end{figure*}

When used to predict mobility data for Oxfordshire, all classes of mobility models presented in Section~\ref{sec:models} underestimate high-volume trips. As the $2\%$ origin--destination pairs with the highest trip volume correspond to $14\%$ of all traffic volume, these trips cannot simply be discarded.

Figure~\ref{fig:raw_models}a shows the overestimation error $T_{ij}^{\text{model}} / T_{ij}^{\text{data}}$ versus the trip volume $T_{ij}^{\text{data}}$, both plotted in logarithmic scale. Larger volume trips are consistently underestimated, having their y-values below the diagonal line marking $T_{ij}^{\text{model}} / T_{ij}^{\text{data}} = 1$.
Considering that the models presented in Section~\ref{sec:models} only take into account the population of different wards and the distance between their centroids, it is not surprising that they ignore the details that make Thames Ward different from Drayton Ward, which lie not in their aggregate demographics, but in their composition in terms of points of interest. Still, one might suggest these differences could be addressed by incorporating measures of accessibility such as travel time\cite{scheurer2007accessibility, geurs2001accessibility}. The reason for this is straightforward: for short trips, travel distance and duration do not scale linearly (see Appendix~\ref{sec:distance}), and thus distance is not always a good proxy for accessibility. One could also try a more straightforward modification to the models, by adding a multiplicative factor to modify the predicted volume for shorter trips. Drawing inspiration from the $e^{-s_{ij}}$ terms in Stouffer's implementation of the intervening opportunities model, and considering that the number of intervening opportunities  $s_{ij}$ grows with distance and travel time, we propose multiplying $T_{ij}$ by a term depending on $s_{ij}$ only, in such a way that large $s_{ij}$ are not penalised, but low $s_{ij}$ lead to an increase in $T_{ij}$. With this extra term, which includes free parameters to be fit by the data, we write a modified expression for the trip volume $T_{ij}$:

\begin{equation}\label{eq:Tij_mod}
T_{ij}^{\text{mod}} = T_{ij} \times \left(1 + A e^{-s_{ij}/b} \right)
\end{equation}

These two possible model modifications, i.e.\ using travel time in place of distance and having the extra dependency on $s_{ij}$, when considered separately or jointly, make a total of three additional models in addition to the original unmodified $T_{ij}$. These three possible modifications, when combined with the eight types of mobility models presented above, give a total of 32 possible mobility models: the eight original models, plus twenty-four possible modifications (three per model).

\begin{figure*}[htp]
\begin{center}
\includegraphics[width=1.0\linewidth]{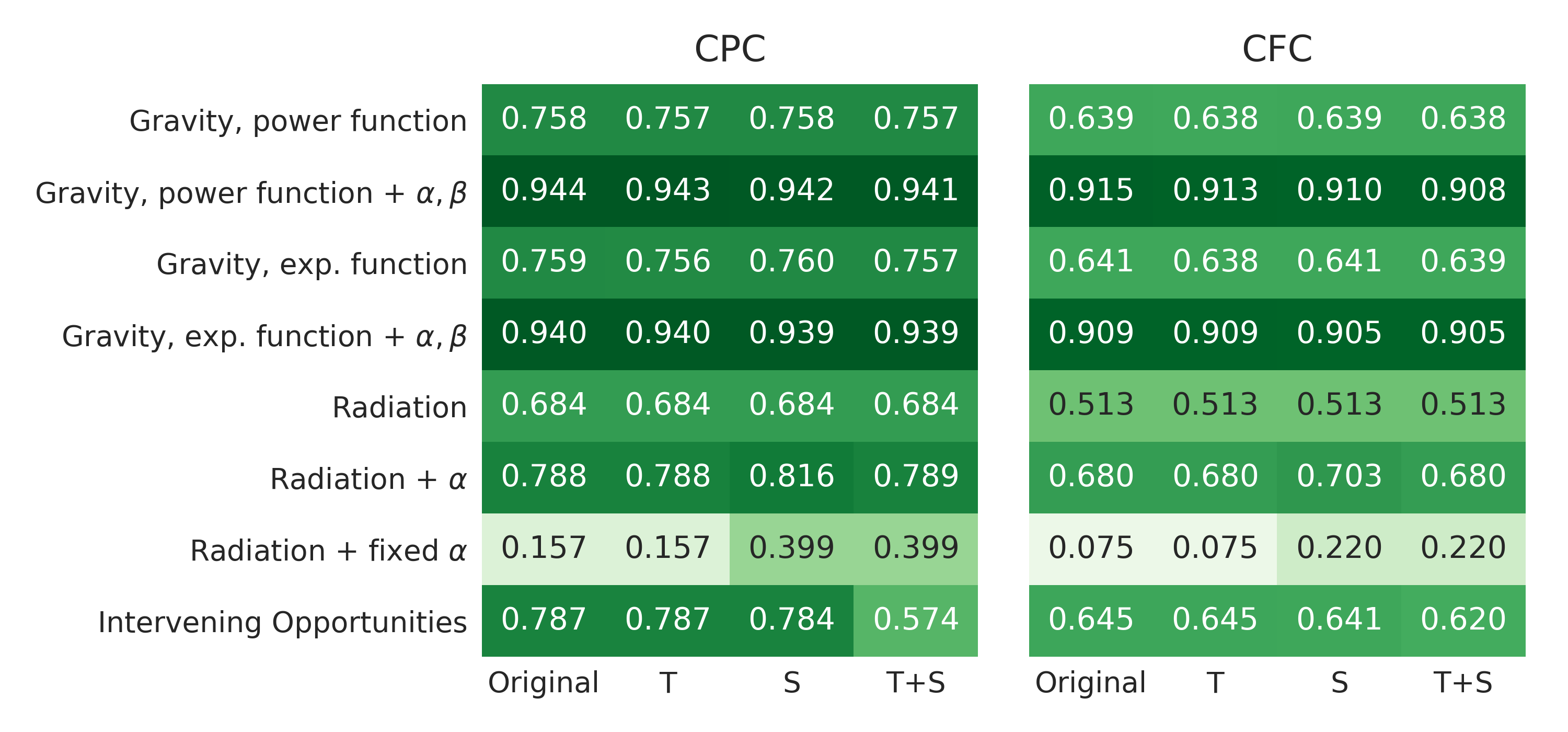}
\end{center}
\caption{\label{fig:heatmap}
\textbf{Simple modifications do not improve model performance.} The two heat maps in this figure show the goodness of fit of different models, as indicated by the CPC and CFC measures. Each panel includes all combinations of the eight models against the four modifications, including the non-modified models. Columns labelled as \texttt{Original}, \texttt{T}, \texttt{S} and \texttt{T+S} respectively represent the original models, followed by models fit using travel time rather than travel distance, models including the $s_{ij}$ modification, and models including both modifications.
$95\%$ confidence intervals were generated from $10,000$ bootstrap samples, resulting in values $< 10^{-4}$ for all estimates.}
\end{figure*}

We calibrated all model parameters using the mean squared error between $\log T_{ij}$ from the model predictions and from the mobility data, using methods from the Python scipy package \cite{jones2014scipy}.
Here we assess the performance of different models using two other traditional goodness-of-fit measures for human mobility models. The first is the common part of commuters (CPC) \cite{gargiulo2012commuting,lenormand2012universal}, based on the S{\o}rensen index \cite{sorensen1948method}:

\begin{equation}\label{eq:CPC1}
\text{CPC}(T^{\text{model}},T^{\text{data}}) \:=\: \frac{ 2 \sum_{i,j} \min(T^{\text{model}}_{ij},T^{\text{data}}_{ij}) }{\sum_{i,j} T^{\text{model}}_{ij} + \sum_{i,j} T^{\text{data}}_{ij}}
\end{equation}

When the total number of commuters $N$ is preserved, the expression for the CPC can be written as a function that decreases with the absolute difference between $T^{\text{model}}$ and $T^{\text{data}}$:

\begin{equation}\label{eq:CPC}
\text{CPC}(T^{\text{model}},T^{\text{data}}) \:=\: 1 - \frac{1}{2N} \sum_{i,j} | T^{\text{model}}_{ij} - T^{\text{data}}_{ij} |
\end{equation}

We also define the Common Fraction of Commuters (CFC), a measure analogous to CPC, but for the ratio between $T^{\text{model}}_{ij}$ and $T^{\text{data}}_{ij}$, in a way that higher values of $T_{ij}$ do not skew the sum, and every $(i,j)$ pair counts equally, regardless of the traffic volume on that origin--destination pair.

\begin{equation}
\text{CFC}(T^{\text{model}},T^{\text{data}}) \:=\: \frac{1}{N} \sum_{i,j} \min \left( \frac{T^{\text{model}}_{ij}}{T^{\text{data}}_{ij}},\frac{T^{\text{data}}_{ij}}{T^{\text{model}}_{ij}} \right)
\end{equation}

The expression for the CPC can be written as a function that decreases with the absolute difference between $\log T^{\text{model}}_{ij}$ and $\log T^{\text{data}}_{ij}$. We provide the full derivation in the Appendix~\ref{sec:CFC}, and the final expression here:

\begin{equation}\label{eq:CPC0}
\text{CFC}(T^{\text{model}},T^{\text{data}}) \:=\:
\frac{1}{N} \sum_{i,j} \exp \left(\: -\frac{1}{2} \left| \log T^{\text{data}}_{ij} - \log T^{\text{model}}_{ij}  \right| \: \right)
\end{equation}

The comparison of how well the 32 models described above predict trips in Oxfordshire can be seen in Figure~\ref{fig:heatmap}. The left and right panels show values for CPC and CFC respectively, for all combinations of the eight models against the three modifications, plus the unmodified models. Each row represents one of the model classes presented in Section~\ref{sec:models}, and columns labelled with \texttt{Original}, \texttt{T}, \texttt{S} and \texttt{T+S} respectively represent the original models, followed by models fit using travel time rather than distance, models including the $s_{ij}$ modification, and models including both modifications.

The first result to be noticed from this figure is that both CPC and CFC produce very similar heat maps. Even though the two measures might differ in their individual values, they follow an approximately linear relation where $\text{CFC} \approx 1.018 \times \text{CPC} - 0.104$, as shown in Appendix~\ref{sec:scalingCPC}. This difference between CPC and CFC is likely to be due to differences in the predicted and observed $T_{ij}$ values for all other trips, which add little to CPC, but add (negative) weight to CFC. Still, according to both measures, the best fit is obtained by the gravity models with unconstrained exponents $\alpha$ and $\beta$ on their population variables $n_i$ and $n_j$.

Besides the difference between both goodness of fit metrics, one can also compare different columns within the same heatmap. There is almost no difference between the non-modified models (\texttt{Original}) and the models modified with travel time (\texttt{T}). 

Adding the $s_{ij}$ modification alone (\texttt{S}), on the other hand, typically makes model predictions worse, except for the modified radiation model with $\alpha$ constrained, where it produces a large improvement in both CPC and CFC. We also generated $95\%$ confidence intervals were from $10,000$ bootstrap samples for all estimates, and all intervals were $< 10^{-4}$, and therefore are not shown in Figure~\ref{fig:heatmap}.
When both modifications are added to the model (\texttt{T+S}), for six out of eight models, the outcome is a worse fit, whereas for two models there is improvement between the non-modified (\texttt{Original}) and the doubly-modified (\texttt{T+S}) cases. Also interesting is the difference between (\texttt{S}) and (\texttt{T}), where adding travel time decreases CPC, while not affecting CFC, for the same six models. This again suggests that swapping travel distance by travel time might not have any clearly positive effect on trip prediction in our setting.

Overall, Figure~\ref{fig:heatmap} shows that replacing the distance between the centroids of each ward by the travel time between them or trying to artificially increase the predicted trip volume for short trips by modifying the functional form of human mobility models does not solve the fact that current mobility models do not perform well at this spatial scale, as presented in Figure~\ref{fig:raw_models}. 

\section{Using OSM data to diagnose the problem}\label{sec:OSM}

\begin{figure*}[htp]
\begin{center}
\includegraphics[width=\linewidth]{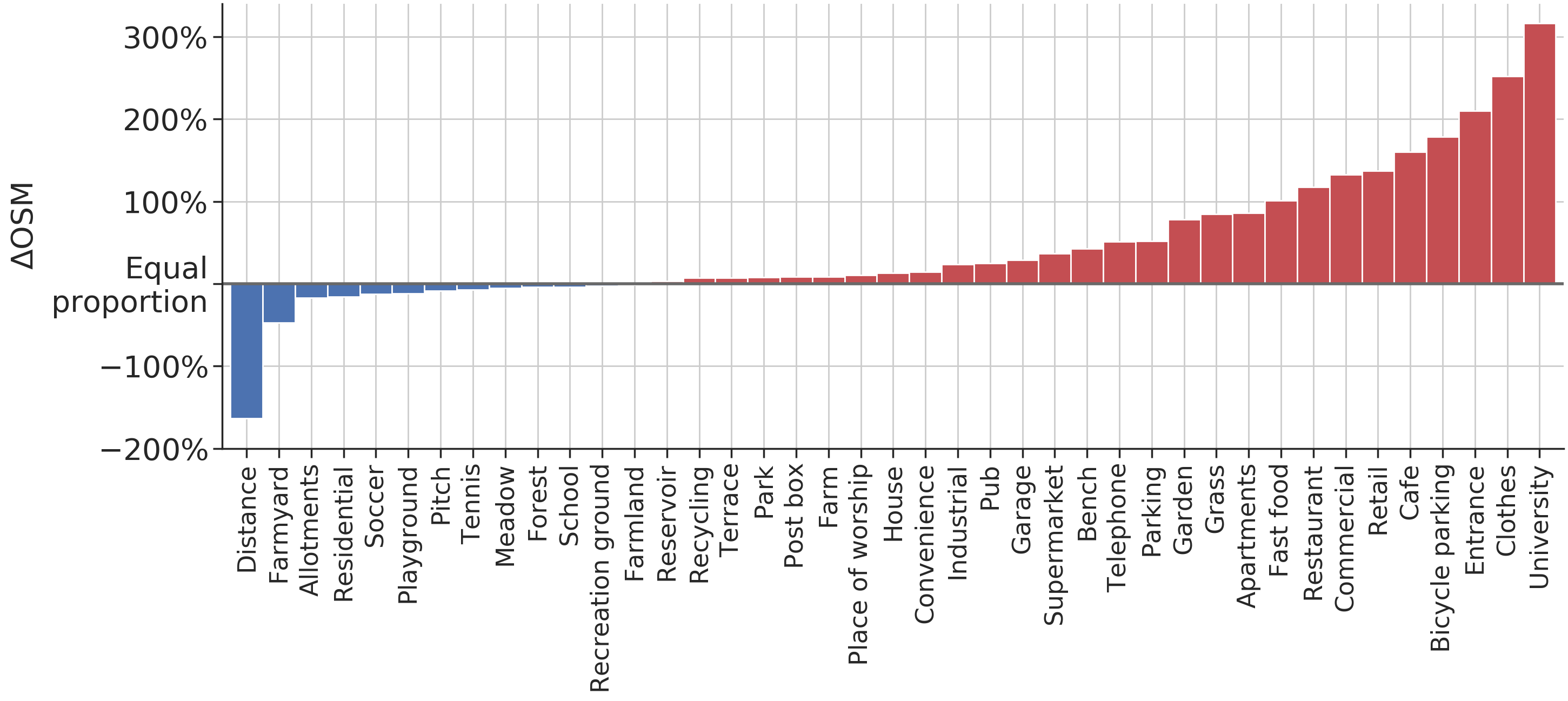}
\caption{\label{fig:OSMbarplot}
\textbf{The average high-volume trip is a short-distance trip between urban areas.} This bar plot shows the difference between the $2\%$ trips with the highest volume and the other $98\%$ of trips. 
It shows all trip attributes (travel distance, or the density of an OSM tag at the destination) which vary over more than $1\%$ between both groups of trips. The length of a bar shows the relative difference between a given trip attribute, when averaged over all high-volume trips, and the same number average over all other trips. For example, the density of points of interest tagged as \cat{fast food} and \cat{restaurant} is approximately $100\%$ higher for high-volume trips, while the average travel distance is $150\%$ lower in the same group.
}
\end{center}
\end{figure*}

The first step towards identifying where current mobility models fail is to observe their prediction accuracy is considerably different for the trips of highest volume, which lie on the right side of every panel in Figure~\ref{fig:raw_models}. In the eight panels, the red dashed line separates the top $2\%$ trips with the highest volume from the other $98\%$. Having split our dataset into two groups, we can then investigate whether the trips in both groups differ in any other attributes, such as their average travel distance, number of intervening opportunities ($s_{ij}$), or the density of any point of interest present in OpenStreetMap. This is a way of addressing the difference illustrated by the trips to Drayton Ward and Thames Ward, as presented in Figure~\ref{fig:wards_illustration}, where two origin--destination pairs of comparable distance and $s_{ij}$ actually led to wards of very different composition in terms of OpenStreetMap points of interest.

Figure~\ref{fig:OSMbarplot} compares the composition of destination wards between the top $2\%$ high-volume trips and all other trips\footnote{Tags corresponding to origin wards have lower values, but show a similar trend.}. 
The figure shows the average distance travelled between wards, as well as the average density of a series of OSM tags in the destination ward, for all OSM categories whose density varies more than $1\%$ between both groups of trips.
In this plot, the length of a bar shows the relative difference between the average value of a trip attribute for all high-volume trips from the average value for all other trips. For example, the number of points of interest tagged as \cat{clothes} is $250\%$ higher for the destinations of high-volume trips and the average distance of high-volume trips is approximately $150\%$ lower compared to all other trips. The figure shows that high-volume trips are often short-distance trips between urban centres. 

\begin{figure*}[htp]
\begin{center}
\includegraphics[width=0.9\linewidth]{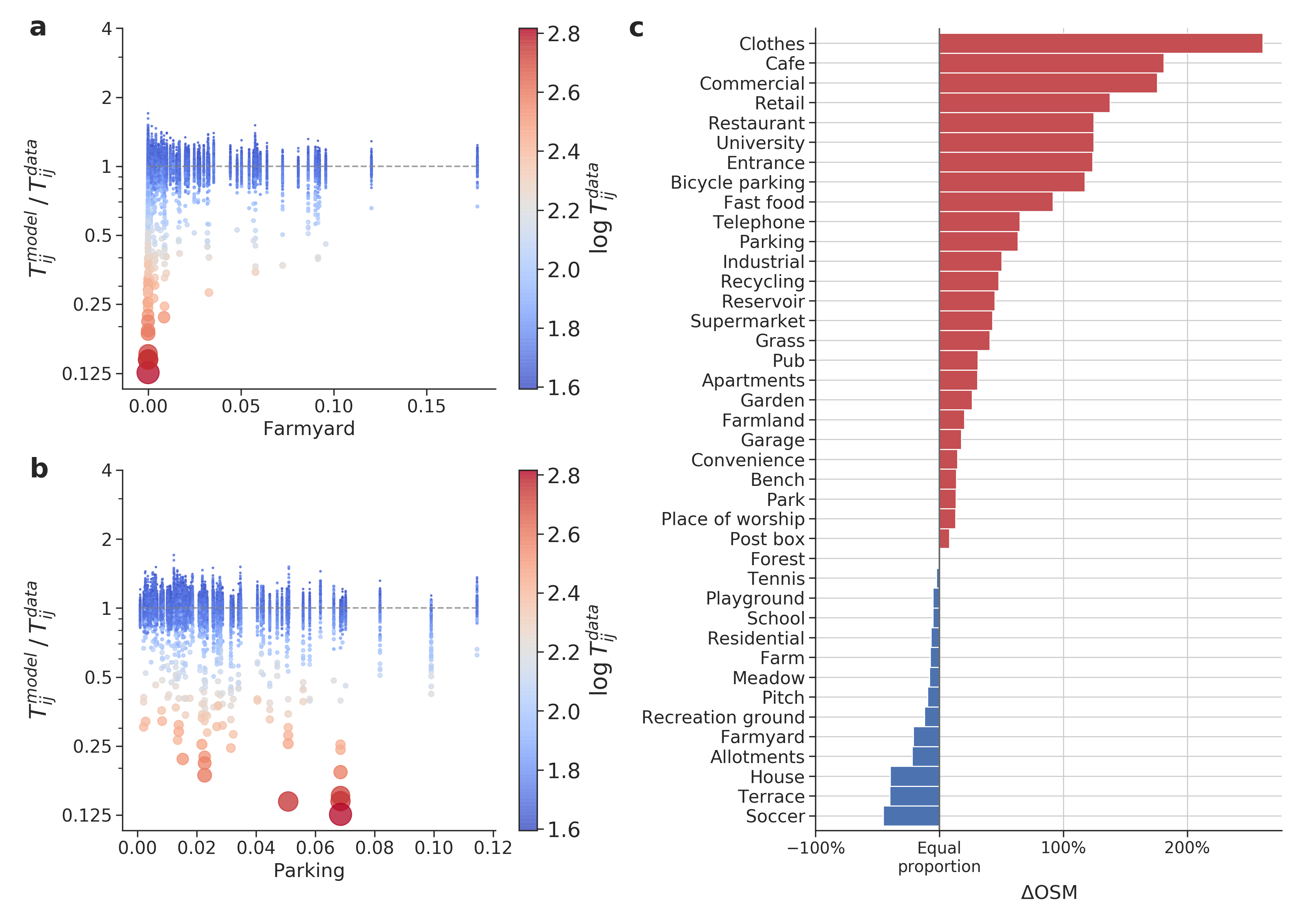}
\caption{\label{fig:overestimation}
\textbf{High-volume urban trips are underestimated by human mobility models.} Panels \textbf{(a)} and \textbf{(b)} show scatter plots for the ratio $T_{ij}^{\text{model}} / T_{ij}^{\text{data}}$, which indicates the overestimation error for the model class with the best performance, namely the gravity model with power functions on the travel cost and unconstrained fitting parameters, against the density of the OSM tags \cat{farmyard} and \cat{parking} in the destination ward. Dot size and colour indicate the magnitude of the real trip volume $T_{ij}^{\text{data}}$ from ward $i$ to ward $j$, with small blue dots for low volumes and large red dots for large volumes. The bar plot in panel \textbf{(c)} shows the difference between the $2\%$ trips with the lowest overestimation ratio when compared to the other $98\%$ of trips. For example, the density of POIs tagged as \cat{cafe} is nearly $200\%$ higher for the underestimated trips, while the density of points POIs tagged as \cat{soccer} is approximately $50\%$ higher for the non-overestimated trips.}
\end{center}
\end{figure*}

Figure~\ref{fig:overestimation} shows how OpenStreetMap data can be used to reveal where human mobility models fail. Figures~\ref{fig:overestimation}a and~\ref{fig:overestimation}b show scatter plots for the ratio $T_{ij}^{\text{model}} / T_{ij}^{\text{data}}$, which indicates the error made by the model. A ratio greater than one indicates overestimation and lower than one indicates underestimation. In both plots, dot size and colour indicate the magnitude of the real trip volume $T_{ij}^{\text{data}}$, with small blue dots indicating low trip volumes and large red dots indicating large trip volumes. In this figure, $T_{ij}^{\text{model}}$ represents predictions using the model with the highest CPC score present in Figure~\ref{fig:heatmap}, i.e.\ the third gravity model discussed in Section~\ref{sec:models}, with a power function on the distance between wards, and its parameters $\alpha$ and $\beta$ unconstrained, without the (\texttt{T+S}) modifications.

On Figure~\ref{fig:overestimation}a, the overestimation ratio is plotted against the density of the OpenStreetMap minor tag \cat{farmyard} in the destination ward, i.e., the number of occurrences of points of interest with that tag, divided by ward population. Most trips are placed around $T_{ij}^{\text{model}} / T_{ij}^{\text{data}} = 1$, meaning little or no overestimation and cover the whole range of values in the x-axis, corresponding to wards with a varying density of points of interest tagged as \cat{farmyard}. By comparison, the high volume trips, marked by the blue dots, are all placed at the lowest densities of \cat{farmyard} points of interest, and the volumes of these trips are consistently underestimated.

The same dots are present in Figure~\ref{fig:overestimation}b, still showing the gross underestimation of high-volume trips, but with the x-axis describing the density of points of interest tagged as \cat{parking}. In this case, the trend for high-volume trips is the opposite of the trend present in panel a: while most trips are distributed over the whole range of \cat{parking} density, high-volume trips are biased towards the higher end of the range. The comparison between Figures~\ref{fig:overestimation}a and~\ref{fig:overestimation}b suggests that the destination wards in high-volume trips are more likely to have parking lots and less likely to have farmyards.

The analysis present in Figures~\ref{fig:overestimation}a and~\ref{fig:overestimation}b can be repeated for many other OpenStreetMap tags. Figure~\ref{fig:overestimation}c shows the result of producing the same scatter plots for the overestimation as presented in the other panels, but for all the OpenStreetMap tags shown in Figure~\ref{fig:OSMbarplot}.
This bar plot is produced after first splitting trip predictions two groups: one group has the $2\%$ trips with the lowest overestimation ratio, corresponding to trips with $T_{ij}^{\text{model}} / T_{ij}^{\text{data}} < 0.686$, and the other group has the other $98\%$, for trips with an overestimation ratio above the same value.
This produces a similar split to separating the top $2\%$ trips with the highest volume, in an agreement of $94\%$ over all origin--destination pairs.
In this bar plot, as in Figure~\ref{fig:OSMbarplot}, the length of a bar shows the relative difference between the average value of a trip attribute or number of points of interest between the two groups, in this case the $2\%$ most underestimated trips on the left and the rest of the dataset on the right. For example, the density of points of interest tagged as \cat{cafe} is nearly $200\%$ higher for the highly underestimated trips, while the density of points of interest tagged as \cat{soccer} is approximately $50\%$ higher for the non-overestimated trips. As a whole, the bar plot indicates that this version of the unconstrained gravity model underestimates urban trips, characterised by a higher density of points of interest tagged as clothes shops, cafes, and general retail.

\section{Discussion}

In this paper, we presented eight classes of human mobility models, including the gravity, radiation, and intervening opportunity models. After calibrating all models using two months of mobility data obtained from smartphone GPS data in the county of Oxfordshire, UK, we showed that most models give poor predictions of traffic volume at this spatial scale. We also showed that typical modifications applied to mobility models, such as exchanging travel distance for travel time or adding extra parameters to correct for poor performance at small spatial scales, do not generally result in any significant improvement in the model fits, which we measured using the common part of commuters (CPC) and the common fraction of commuters (CFC). 

The poor performance of current mobility models in predicting trips within Oxfordshire is in agreement with the literature, which shows several cases where this problem is successfully addressed by the inclusion of novel data sources to the predictive model. These new data sources typically include data on the accessibility of different regions of a city, including but not limited to road configurations~\cite{akbarzadeh2018communicability}, city block shapes~\cite{brelsford2018toward}, travel cost and number of opportunities~\cite{Batty2009}, among many variables which provide a more fine-grained description of how regions of a city differ from each other. We believe these approaches are important, as they repeatedly show that describing human mobility at small spatial scales in urban environments requires further differentiating between regions which might otherwise be similar at an aggregate level.

We performed a more thorough diagnosis of where mobility models fail using OpenStreetMap data. This analysis revealed that the trips of highest volume have a specific profile in terms of their density of points of interest. The $2\%$ trips with the highest volume trips have a high density of points labelled as university facilities, clothes and other forms of retail, cafes, restaurants, and other urban landmarks, with densities sometimes over $200\%$ higher than all other trips. High-volume trips are also typically over much shorter distances. The number of intervening opportunities between wards $i$ and $j$, measured by the variable $s_{ij}$, is also very different between the two portions of trips, as $s_{ij}$ is on average over $600\%$ higher in the $98\%$ contingent. While some of the specific features (e.g., university facilities) are likely unique to Oxford, we have no reason to believe the general pattern of models underestimating short distance trips between urban areas is specific to Oxfordshire, which future work should investigate.

We also showed that those high-volume trips are severely underestimated even by the best-performing mobility models, sometimes being predicted at only approximately $12.5\%$ of the actual traffic volume. Taking the the mobility model with the best performance for further analysis, we split trips into the $2\%$ most underpredicted and the remaining $98\%$, and show that the origin--destination pairs for which the model performs worst are indeed the most urban ones, characterised by the same OpenStreetMap tags as the set of trips of highest volume. 

Unlike data sources that might only be available at a specific time or location or that might incur the expense of deploying large numbers of urban sensors, the OpenStreetMap dataset used in this paper is freely available worldwide. As long as there is enough coverage, OpenStreetMap data can be a powerful and easily applicable tool when combined with other demographic and geographic data to diagnose the performance of multiple human mobility models. This is in stark contrast more expensive approaches requiring traffic sensors or opaque methods, which might not be viable options for local governments, for reasons of time, skills, or budgetary constraints. Instead, the diagnosis method we describe here does not require any expensive computation, sensor deployment or proprietary data.

As explained above, this method is flexible and easily applicable to multiple human mobility models, as long as there is enough OpenStreetMap coverage in the region. Who contributes to OpenStreetMap and potential biases of the data are an important related area \cite{thebault2018geographic,thebault2018distance,dasgendered}.

This work also adds to the discussion of how to model human mobility at different spatial scales. While the dataset in this paper describes traffic within the perimeter of a region of roughly 2,605 km$^2$, mobility models have been tested and compared across different scales~\cite{masucci2013gravity,Piovani2018}, including their many successful applications describing cross-country migration~\cite{lenormand2012universal,simini2012universal,mcneill2017estimating,najem2018debye}. At a national scale, many of the modifications discussed here should not be necessary, since travel time scales with travel distance for larger scales (see Appendix~\ref{sec:distance}), but also since large, city-sized regions are likely to contain wider distributions of OpenStreetMap points of interest, rather than the very uneven distributions observed for different Oxfordshire wards.

Given how understanding human mobility at small spatial scales is important for urban planning, it is important to have tools which can flexibly incorporate diverse sources of land use and accessibility data into model diagnosis and traffic prediction. Here we have shown how OpenStreetMap can be one tool, providing useful insights to the study of how people move in urban landscapes.

\section*{Data accessibility}
Code will be available upon acceptance. 

\section*{Funding statement}
 This project was supported by funding from InnovateUK under grant number 52277-393176, the NERC under grant number NE/N00728X/1, and the Lloyd's Register Foundation.

\section*{Competing interests}
The authors declare no competing financial interests.

\section*{Author contributions}
All authors conceived and designed the study and collected the data. CQC implemented the models, carried out the analysis and wrote the first draft. SAH and JB secured the funding and SAH coordinated the project. All authors edited the manuscript and gave final approval for publication.

\appendix

\section{Scaling of measures of travel distance}\label{sec:distance}

\begin{figure*}[htp]
\begin{center}
\includegraphics[width=\linewidth]{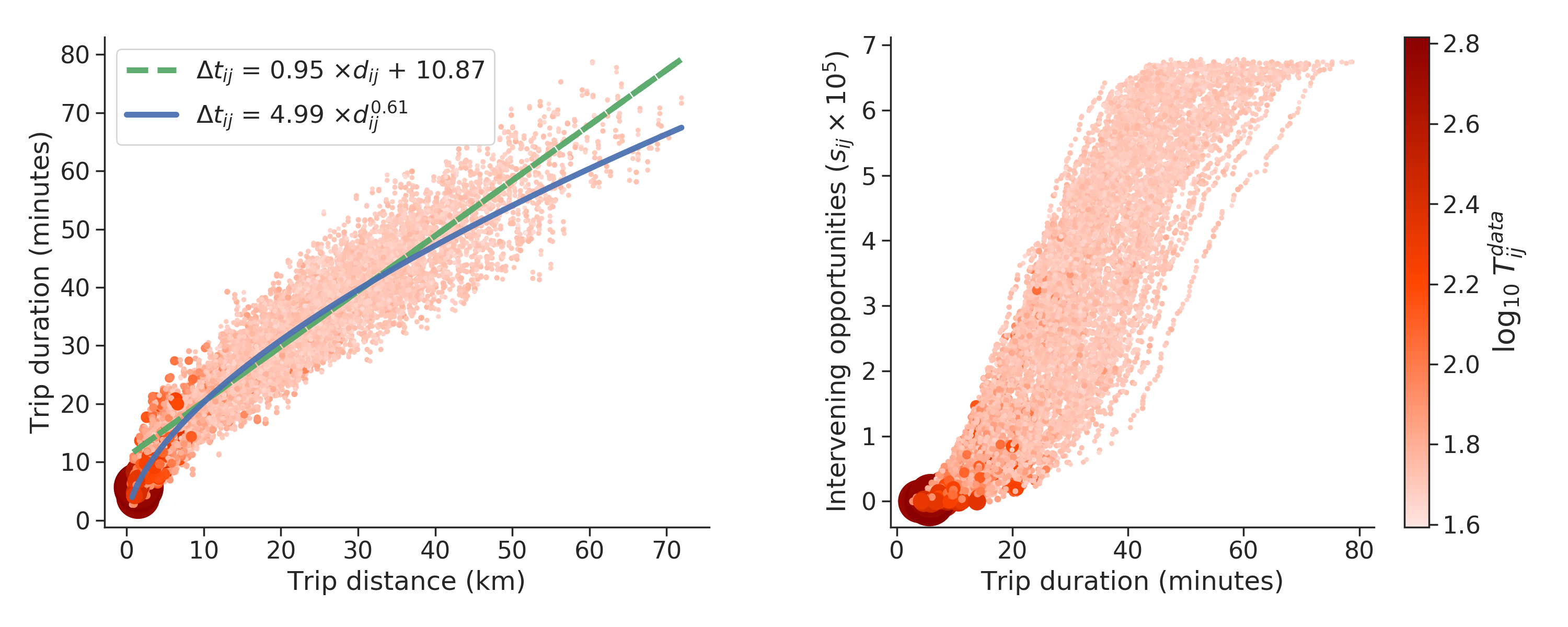}
\caption{\label{fig:distance}
\textbf{Scaling between different metrics for travel distance.} The left panel shows how trip duration measured in minutes and travel distance measured in kilometres scale in an almost linear fashion. The right panel shows trip duration on the x-axis, and intervening opportunities ($s_{ij}$) on the y-axis. The plot shows a clear positive correlation between both variables ($r = 0.86, p < 10^{-16}$). In both panels, smaller lighter dots indicate trips with a lower trip volume, whilst larger darker dots indicate higher trip volumes.
}
\end{center}
\end{figure*}

Figure~\ref{fig:distance} shows how different metrics for travel distance scale with each other. The left panel shows how trip duration in minutes and the distance in kilometres between wards scales in an almost linear fashion, with a Pearson coefficient of $r = 0.94$ ($p < 10^{-16}$) differing only for trips shorter than 10km or less than 15 minutes driving. Two fits are also shown on the figure: a linear fit showing trip duration can be approximated by $\delta t_{ij} = 0.95 \times d_{ij} + 10.87$, and a nonlinear fit estimating trip duration as $\delta t_{ij} = 4.99 \times d_{ij}^{0.61}$. For both fits, $\delta t_{ij}$ is the trip duration measured in minutes, and $d_{ij}$ is the distance in kilometres between wards $i$ and $j$. The right panel shows trip duration in minutes on the x-axis, and the number of intervening opportunities between wards $i$ and $j$, given by the variable $s_{ij}$, on the y-axis. The plot shows a clear positive correlation between both variables ($r = 0.86, p < 10^{-16}$). In both panels, smaller lighter dots indicate trips with a lower trip volume (i.e., lower $T_{ij}$), whilst larger darker dots indicate higher trip volumes.

\section{Scaling of CPC with CFC}\label{sec:scalingCPC}

\begin{figure*}[htp]
\begin{center}
\includegraphics[width=0.6\linewidth]{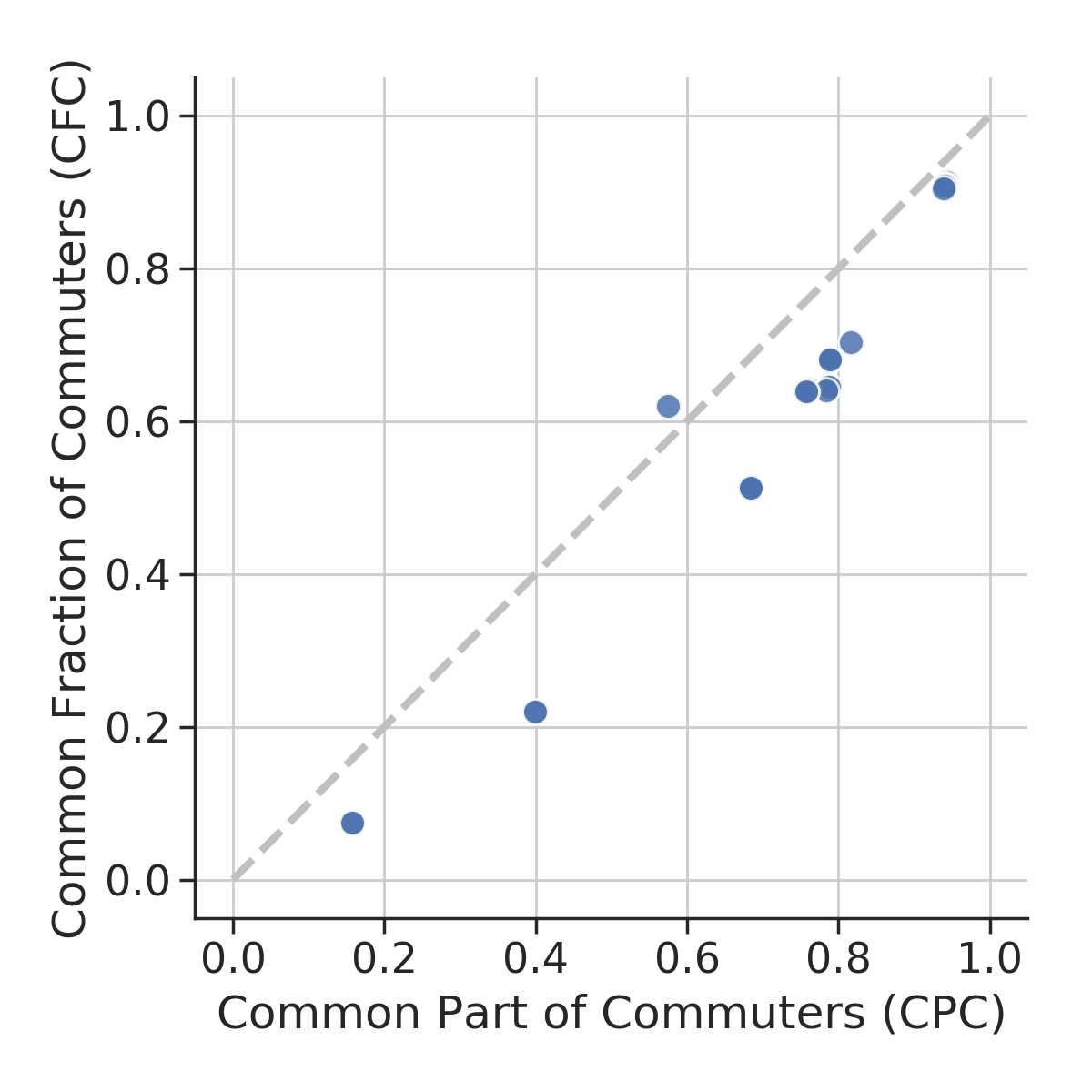}
\caption{\label{fig:scalingCPC}
Scaling between the Common Part of Commuters (CPC) and the Common Fraction of Commuters (CFC) for the model variations shown in Figure~\ref{fig:heatmap}. The variables follow an approximately linear relation, with $\text{CFC} \approx 1.018 \times \text{CPC} - 0.104$.
}
\end{center}
\end{figure*}

Figure~\ref{fig:scalingCPC} shows the scaling between the Common Part of Commuters (CPC) and the Common Fraction of Commuters (CFC) for the model variations shown in Figure~\ref{fig:heatmap}. The variables follow an approximately linear relation, with $\text{CFC} \approx 1.018 \times \text{CPC} - 0.104$.

\section{Derivation of how CFC is the L1 norm on \texorpdfstring{$\log T_{ij}$}{log Tij}}\label{sec:CFC}
The last line uses the fact that $\min(x,y) = (x+y -|x-y| )/2$. See how the last line is a sum of absolute-value differences between $\ln T_0$ and $\ln T_1$, making CFC close to a L1 norm on the log $T_{ij}$.

\begin{eqnarray*}
\frac{1}{N} \sum_{ij} \min \left( \frac{T_0}{T_1}, \frac{T_1}{T_0} \right) 
&=& \frac{1}{N} \sum_{ij} \min \left( e^{\ln T_0 - \ln T_1}, e^{\ln T_1 - \ln T_0} \right) \\
&=& \frac{1}{N} \sum_{ij} \exp \left[ \: \min \left( \ln T_0 - \ln T_1, \ln T_1 - \ln T_0 \right) \: \right] \\
&=& \frac{1}{N} \sum_{ij} \exp \left( \: - \frac{1}{2} \left| \ln T_0 - \ln T_1 \right| \: \right)
\end{eqnarray*}

\section{The threshold for high traffic is robust}\label{sec:threshold}

\begin{figure*}[htp]
\begin{center}
\includegraphics[width=\linewidth]{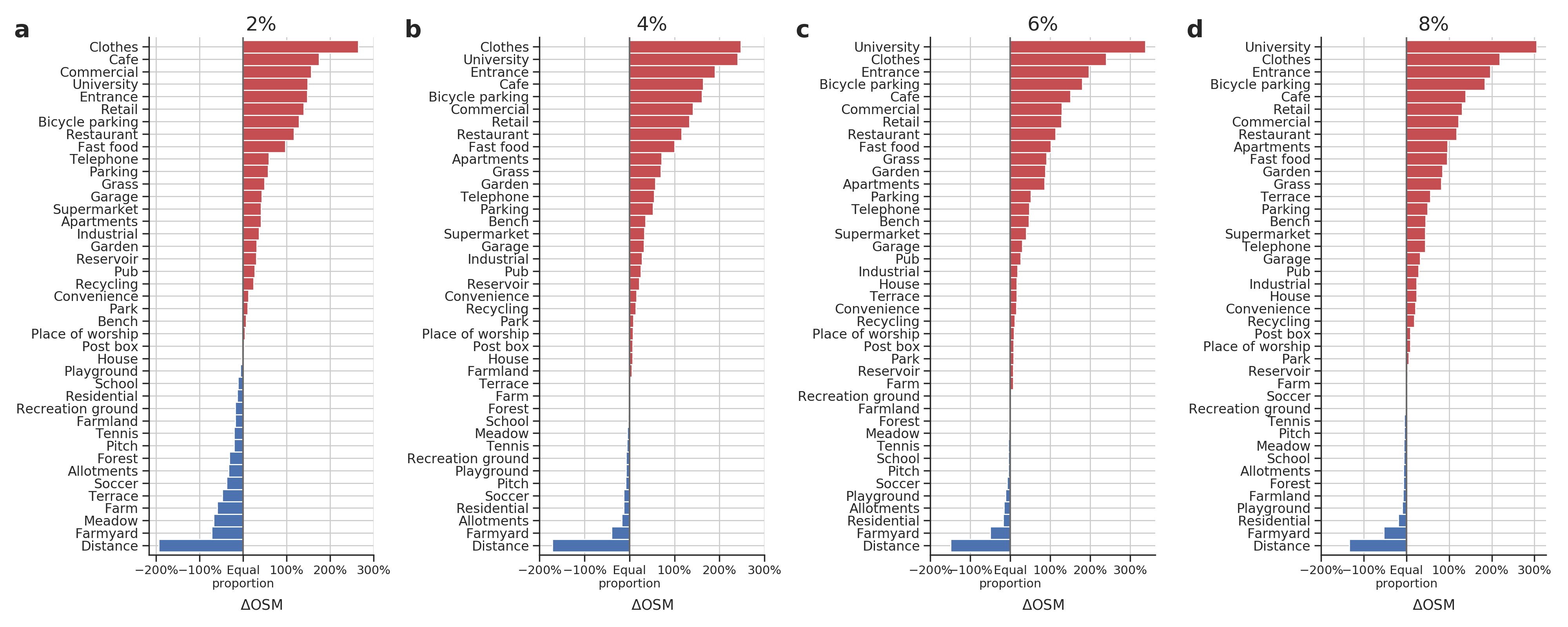}
\caption{\label{fig:threshold}
\textbf{The high-volume split is robust to the cutoff threshold.}
Panels \textbf{(a)} to \textbf{(d)} show bar plots similar to Figure~\ref{fig:OSMbarplot}, but for different cutoffs, ranging from $2\%$ to $8\%$. It shows no meaningful variation in the properties of the high-volume trips and the properties of all other trips in terms of OpenStreetMap points of interest. In all cases, high-volume trips are on average over shorter distances and to more urban zones.
}
\end{center}
\end{figure*}

Figure~\ref{fig:threshold} shows a series of bar plots similar to Figure~\ref{fig:OSMbarplot}, but for different high-volume cutoffs, ranging from $2\%$ to $8\%$. It shows no meaningful variation in the properties of the high-volume trips and the properties of all other trips in terms of OpenStreetMap points of interest. In all cases, high-volume trips are on average shorter and go to more urban places than do other trips.

\end{document}